\documentclass[english]{article}
\usepackage[T1]{fontenc}
\usepackage[latin9]{inputenc}
\usepackage{amsmath}
\usepackage{url}

\usepackage{graphics}
\usepackage{graphicx}
\usepackage{epsfig}
\usepackage{float}
\usepackage{xcolor}
\usepackage{afterpage}
\usepackage{placeins}

\usepackage{babel}

\makeatletter

\providecommand{\tabularnewline}{\\}

\makeatother

\usepackage{babel}
\begin{document}

\title{Bose-Hubbard model with a single qubit}

\author{R. M. Woloshyn\\
TRIUMF, 4004 Wesbrook Mall\\
Vancouver, British Columbia, V6T 2A3, Canada}

 
\maketitle
\begin{abstract}
The use of a single-qubit parametrized circuit as an \emph{Ansatz} for the
variational wave function in the calculation of the ground state energy
of a quantum many-body system is demonstrated using the one-dimensional
Bose-Hubbard model. Comparison is made to calculations where a classic
neural network is used to generate the variational wave function.
Computations carried out on IBM Quantum hardware are also presented.
\end{abstract}

\section{Introduction}

Machine learning based methods are becoming more prevalent in physics.
One idea that has gained considerable traction is the use of a neural
network as a variational \emph{Ansatz} for many-body systems 
\cite{doi:10.1126/science.aag2302} with applications in condensed matter 
physics \cite{PhysRevB.96.205152,Choo2020,Yoshioka2021}
and nuclear physics \cite{KEEBLE2020135743,PhysRevLett.127.022502,Gnech2021}. 
In particular, neural quantum states were used in \cite{saito2017, saitokato}
to determine the ground state of
the Bose-Hubbard model. Zhu \emph{et al.} \cite{PhysRevResearch.5.043084}
extended these calculations in a way that allows for determining ground 
and excited states over a wide range of parameters with only minimal training.

Another recent trend is the exploration of quantum computing for physics
research \cite{garcíaramos2023nuclear,dimeglio2023quantum}. 
Relevant here is the synergy between machine learning and
quantum computation. In general, where a classical neural network is being used,
one may consider replacing it by a quantum circuit. In this note
the use of a parametrized quantum circuit in lieu of the classic neural
networks of Refs. \cite{saito2017, saitokato,PhysRevResearch.5.043084}
for the Bose-Hubbard model is discussed. 

In Ref. \cite{PerezSalinas2020datareuploading}, P\'erez-Salinas \emph{et al.} 
proposed a universal quantum classifier, which, with data multiplexing
and re-uploading \cite{vidal2020input}, uses only one qubit. This idea was motivated 
by the observation that in a classical neural network all features are input into
all nodes while in typical quantum neural networks \cite{kwak2021quantum,zhao2021review}
each feature is encoded and first processed by a separate qubit. Using 
data re-uploading with a single qubit, loading the features in a way 
more analogous to that of a neural network could be achieved. In a subsequent 
paper \cite{PhysRevA.104.012405},  this notion was formalized in the Quantum UAT, a quantum version of the 
Universal Approximation Theorem \cite{Cybenko1989,HORNIK1991251} for classical
neural networks.
A circuit with two parametrized rotation gates acting on a single qubit for
implementing the Quantum UAT was presented. In
this work, the circuits of \cite{PerezSalinas2020datareuploading,PhysRevA.104.012405}
are not used as classifiers as originally proposed, but as variational 
\emph{Ansatze} which are trained to generate the ground state
wave function of the one-dimensional Bose-Hubbard model.

Sect. 2 introduces the Bose-Hubbard model and, for comparison to subsequent
quantum computations, a variational calculation using a classical
neural network for the wave function is presented. The calculation
is done in the Fock state basis for a model with six sites and five
bosons. This matches the one-dimensional calculation done by Zhu \emph{et al.} 
\cite{PhysRevResearch.5.043084}.
In this section we also discuss how to use the symmetries of
the ground state to reduce the size of the required basis to something
more suitable for the quantum computation.

Quantum computations, using a classical computer to simulate an ideal
quantum device, are presented in Sect. 3. Two different circuits are
considered: the compressed scheme (Fig. 2(b) in \cite{PerezSalinas2020datareuploading})
and the universal quantum approximant from \cite{PhysRevA.104.012405}. 
Calculations are carried out at a
few benchmark points. The results clearly show convergence toward
the exact ground state energies but with an accuracy which is not
quite as good as the classical neural network. However, the quantum
simulations contain a very small number of parameters compared to
the neural network and, to save time, used fewer update steps. 

To get some insight into the effect of hardware noise on the single
qubit calculations energies were calculated on the IBM Quantum processor
ibm\_brisbane. The results, without any error mitigation,
are presented in Sect. 4

The wave function of the Bose-Hubbard model is real but, in general,
one may have to deal with a complex wave function. The Appendix demonstrates
how to do this using a deformation of the Bose-Hubbard Hamiltonian with
a complex phase in the hopping term.

\section{Bose-Hubbard model}

The Hamiltonian for the Bose-Hubbard model, in the absence of an external
and chemical potential, is 

\begin{equation}
\mathcal{H}=-t\sum_{<i,j>}a_{i}^{\dagger}a_{j}+\frac{U}{2}\sum_{i}^{M}n_{i}(n_{i}-1),
\end{equation}
where $a_{i}^{\dagger}$and $a_{i}$ are the usual boson creation
and annihilation operators with the commutation relation $\left[a_{i},a_{j}\right]=\delta_{ij}$
and $n_{i}=a_{i}^{\dagger}a_{i}$ is the number operator at site $i.$
Periodic boundary conditions are imposed.

Calculations are typically done in the basis of Fock states which
enumerate the number of bosons at each site. For $N$ bosons and $M$
sites the total number of Fock states is 
\[
\dbinom{N+M-1}{N}.
\]
The basis size grows exponentially so, for example, in \cite{saitokato}
with 8 sites, the basis is too large for an exact evaluation of the energy.
Monte Carlo sampling was used to approximate the expectation value
of the Hamiltonian. For the present calculation, 
following \cite{PhysRevResearch.5.043084} with,
with $M=6$ and $N=5$ the number of Fock states is only 252 so the
full basis can be used. 

The ground state energy is calculated using a variational method and a
neural network is used to determine the coefficients of a Fock state
expansion of the wave function \cite{saitokato,PhysRevResearch.5.043084}. 
The network parameters are the variational 
parameters that are tuned to minimize the energy. In machine learning 
terminology, the site occupation numbers of the Fock states are the features and 
Fock states comprise the data sample. Note that unlike a typical machine
learning classification problem, mini-batches are not used here. 

The neural network used here is comprised an input layer with 6 inputs
corresponding to the number of sites, two hidden layers with 64 and
32 nodes respectively and a linear output layer with a single output.
Each hidden layer has a Tanh activation. The total number of parameters
is 2560. Following \cite{saitokato}, 
the network input is pre-processed by subtracting
the average occupancy $N/M$ from the individual Fock state occupation
values. The wave function coefficients are calculated by taking the
exponential of the network output. This ensures that all coefficients
have the same sign as in the true wave function.

The network was implemented using Pytorch \cite{paszke2019pytorch}. 
The Adam optimizer with
learning rate of 0.02 was used. Training runs consisted of 1500 update
steps using the full sample of 252 basis states. Results are shown
in Table 1 along with the results from exact diagonalization.

\begin{table}
\begin{centering}
\begin{tabular}{|c|c|c|c|}
\hline 
  & & Neural network & Neural network\tabularnewline
\raisebox{.8ex}{U} & \raisebox{.8ex}{Exact} & full basis & reduced basis\tabularnewline
\hline 
2 & -7.54752 & -7.54750 & -7.54747\tabularnewline
\hline 
5 & -5.46241 & -5.46236 & -5.46233\tabularnewline
\hline 
8 & -4.37439 & -4.37439 & -4.37388\tabularnewline
\hline 
\end{tabular}
\caption{Results for the ground state energy for different values of the on-site
interaction energy $U$ at $t=1$ using a neural network variational
\emph{Ansatz}. Results of exact diagonalization are also shown.}
\par\end{centering}
\end{table}

The full Fock state wave function contains many redundant values.
Due to translation invariance with periodic boundary conditions, Fock
states whose occupation numbers are cyclic permutations of each other
have the same weight. By using a basis of composite states consisting
of the 6 states related by translation the basis size can be reduced
from 252 to 42. Furthermore, in the ground state, Fock states whose
occupation numbers are related by a parity transformation, for example,
$\left|012011\right\rangle $ and $\left|110210\right\rangle ,$ have
the same coefficient. This can be used to reduce the the basis size,
from 42 to 26 in the present case. Neural network variational results
using the 26 states basis are given in the last column of Table 1. This
is the basis that will be used in the quantum computation. Note that
it can only be use to get ground state since in excited states not
all wave function coefficients have the same sign so the composite
states used here will not be applicable. 

\section{Quantum computation with a single qubit}

A variety of quantum counterparts to classical neural networks have
been proposed. For a review see \cite{kwak2021quantum,zhao2021review}. 
In the typical architecture,
each feature is loaded in a separate qubit and processing follows.
The circuits used here differ in two essential ways. First, multiple
data are loaded in the same input channel. Secondly, loading and processing
the data take place in the same quantum gates. 

\begin{figure}[tb]
\centering
\includegraphics[scale=0.4,trim={5.cm 0cm 4.8cm 0cm},clip=true,angle=270]{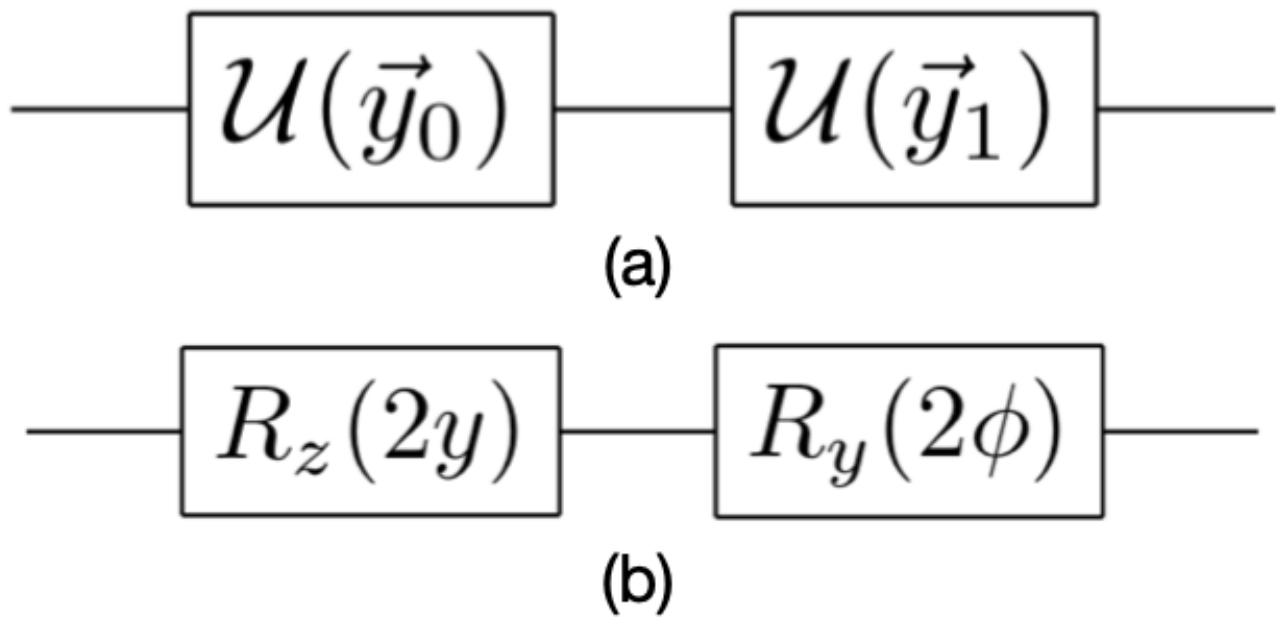 }
\caption{One-qubit circuits: (a) One layer of the compressed scheme circuit. The
gate $\mathcal{U}$ is a general single-qubit unitary. 
The 6 features $\vec{x}$  are divided into 2 triplets. There are 7 parameters per layer, 
6 weights $\vec{\omega}$ and 1 bias $b$, the same for all triplets. 
(b) Implementation of the fundamental UAT gate of \cite{PhysRevA.104.012405}.
There are 8 parameters per layer, 6 weights $\vec{\omega}$
and 1 bias $b$ plus a rotation angle $\phi$.}
\label{oqc}
\end{figure}

The two different circuits used in this work are shown in Fig. \ref{oqc}. 
Fig. {\ref{oqc}}(a) is based on the universal quantum classifier using the compressed
scheme proposed in \cite{PerezSalinas2020datareuploading}. 
The input data are grouped in triplets
$\vec{x}$ and combined with variational parameters $(\vec{\omega},b)$
to form the arguments of general unitary operators. In the present
application, with six input features, a layer of this circuit has
two unitary operators with arguments given by

\begin{equation}
\vec{y}_{j} = [b+{\omega}_{3j}x_{3j}, b+{\omega}_{3j+1}x_{3j+1}, b+{\omega}_{3j+2}x_{3j+2}], j={0,1.}
\end{equation}
Data are re-uploaded using multiple layers of the basic circuit with
each layer having seven parameters.

An alternative circuit QUAT, uses multiple layers of the fundamental UAT 
gate \cite{PhysRevA.104.012405} shown in Fig. {\ref{oqc}}(b). 
All features $\vec{X}$ are combined into a single variable given by 

\begin{equation}
y=\vec{\omega}\cdot\vec{X}+b,
\end{equation}
The rotation gate $R_{y}(2\phi)$ is meant to play a role like the activation
function of a neural network. With six input features there are eight
parameters per layer.

The reduced 26-state basis is used in the quantum calculations. A
single Fock state was chosen from each of 26 classes making up the
composite basis states to serve as a representative input sample.
For each of these, the average value of the qubit in the computational 
basis
after application of the circuit is used as the weight of the corresponding 
basis state in the variational wave function. The approach here is a hybrid
one, the calculation of wave function weights is quantum, the energy
calculation and minimization are classical.

To determine the efficacy of the single qubit circuits as variational
\emph{Ansatze} computations were done on a classical computer simulating
an ideal quantum computer. In other words, there was no quantum measurement
uncertainty and no hardware noise. These issues will be discussed
briefly in the next section. The Pennylane quantum programming framework
\cite{bergholm2022pennylane}
was used to implement the simulation. The Adam optimizer, running on a 
classical computer,  was used for energy minimization.

The first step is to determine if the single qubit \emph{Ansatz} will actually
work and how many layers may be required. The loss, defined as the
exact ground state energy minus the variational estimate, as a function
of upgrade step using different numbers of the compressed scheme circuits
for the model parameters $t=1$ and $U=5$. The number of parameters
increases from 21 with 3 layers to 42 with 6 layers. Table 2 shows how
the estimated energy varies with number of layers for both types of
circuits.

\begin{table}
\begin{centering}
\begin{tabular}{|c|c|c|}
\hline 
Layers & Compressed scheme & QUAT circuit\tabularnewline
\hline 
3 & -5.40019 & -5.02048\tabularnewline
\hline 
4 & -5.45876 & -5.35662\tabularnewline
\hline 
5 & -5.45396 & -5.39576\tabularnewline
\hline 
6 & -5.46048 & -5.46175\tabularnewline
\hline 
\end{tabular}
\par\end{centering}
\caption{Variational energy estimate after 1200 training steps
with model parameters $t=1$ and $U=5$ for different numbers of layers of
the basic circuits in Fig. \ref{oqc}. }
\end{table}

\begin{table}
\begin{centering}
\begin{tabular}{|c|c|c|c|c|}
\hline 
  & & Neural network & Quantum  & Quantum \tabularnewline
\raisebox{.8ex}{U} & \raisebox{.8ex}{Exact} & reduced basis & compressed & QUAT\tabularnewline
\hline 
2 & -7.54752 & -7.54747 & -7.54750 & -7.54751\tabularnewline
\hline 
5 & -5.46241 & -5.46233 & -5.46048 & -5.46175\tabularnewline
\hline 
8 & -4.37439 & -4.37388 & -4.36807 & -4.31310\tabularnewline
\hline 
\end{tabular}
\par\end{centering}
\caption{Comparison of the variational energy estimate from quantum simulation
with exact and neural network values at $t=1$ and different values of $U$. 
The quantum simulations used 6 layers and were terminated after 1200 update steps.}
\end{table}

Finally, the results of quantum simulation are compared to exact and
neural network values for the three benchmark model parameter sets
in Table 3. The results using a single qubit are very reasonable
although slightly less accurate than the neural network. However, it should
be kept in mind that the quantum circuits have about 50 times
fewer parameters than the neural networks and there may still be room
to improve the results by increasing training and tuning the optimizer.

\section{Noisy simulation}

\begin{figure}[t]
\centering
\includegraphics[scale=0.6,trim={0cm 0cm 0.5cm 0.cm},clip=true,angle=0]{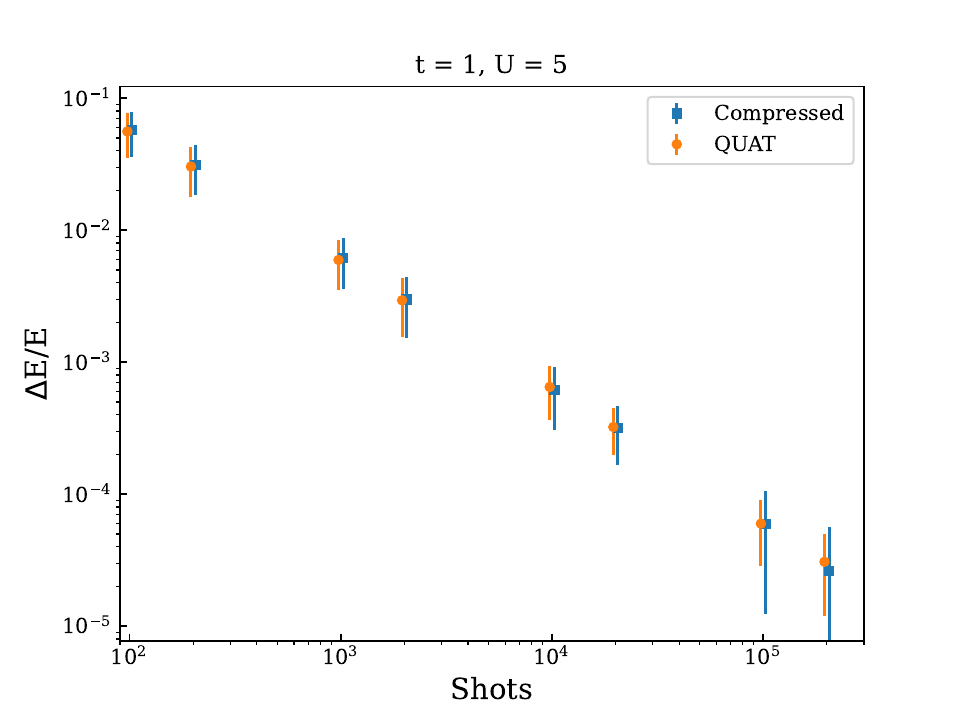 }
\caption{Fractional energy difference between an ideal simulation and a
simulation with finite shots. Symbols are median values from 100 trials. Error
bars reflect the standard deviation of the trial distribution. } 
\label{nshots}
\end{figure}

Since access to quantum resources was limited, it was not feasible
to carry out training an a real quantum device. However, it is possible
to gain some insight into the performance of the circuits by calculating
the energy using trained circuits on a real device. The first step
is to determine how the number of measurements (shots) affects the
calculation of the wave function and hence the value of the energy.
This was done on the quantum simulator without hardware noise. Fig. \ref{nshots}
shows the difference between the energy calculated using a finite number of
shots and the ideal value (essentially infinite shots) as fraction of the 
ideal energy. The results were obtained by taking the median of 100 trials
for each value of the shots. The error bars reflect the standard deviation
of the distribution of trial values. A practical limit for an actual 
quantum computation is about 20,000 shots where $\Delta E/E$ is 
below $10^{-3}.$

\begin{figure}[th]
\centering
\includegraphics[scale=0.6,trim={0cm 0cm 0.5cm 0.cm},clip=true,angle=0]{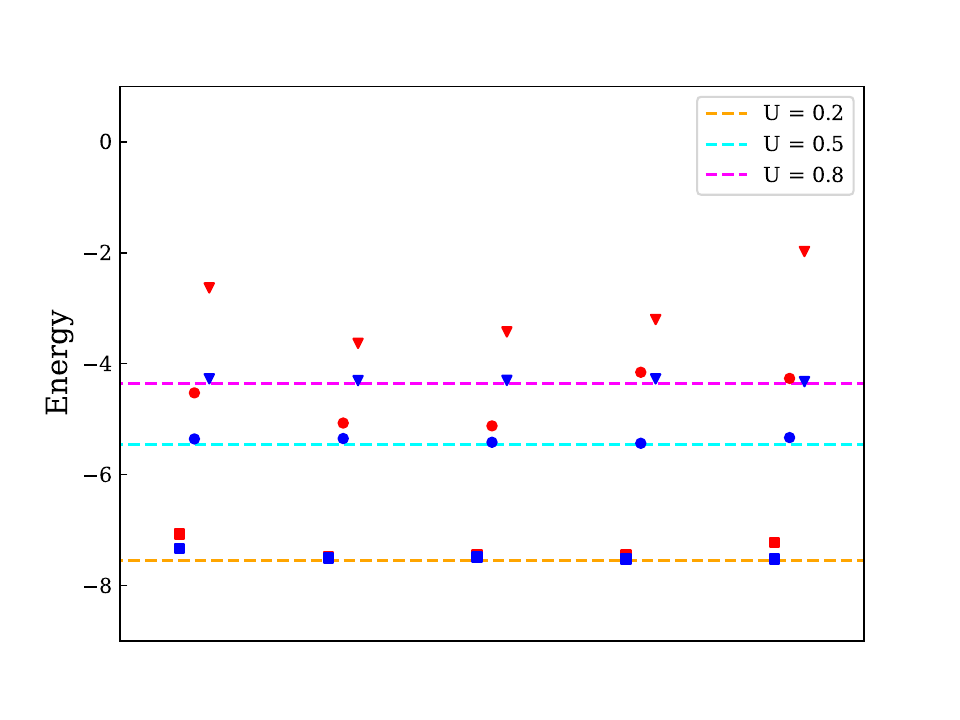 }
\caption{Five ground state energies using wave functions computed in a single
run on ibm\_brisbane with (blue symbols) and without (red symbols) readout error 
correction using the compressed scheme circuit. 
Triangles, circles and squares correspond to 
$U$ = 0.2, 0.5 and 0.8 respectively.
The ideal hardware error-free results are shown by dashed lines } 
\label{plt5c}

\centering
\includegraphics[scale=0.6,trim={0cm 0cm 0.5cm 0.cm},clip=true,angle=0]{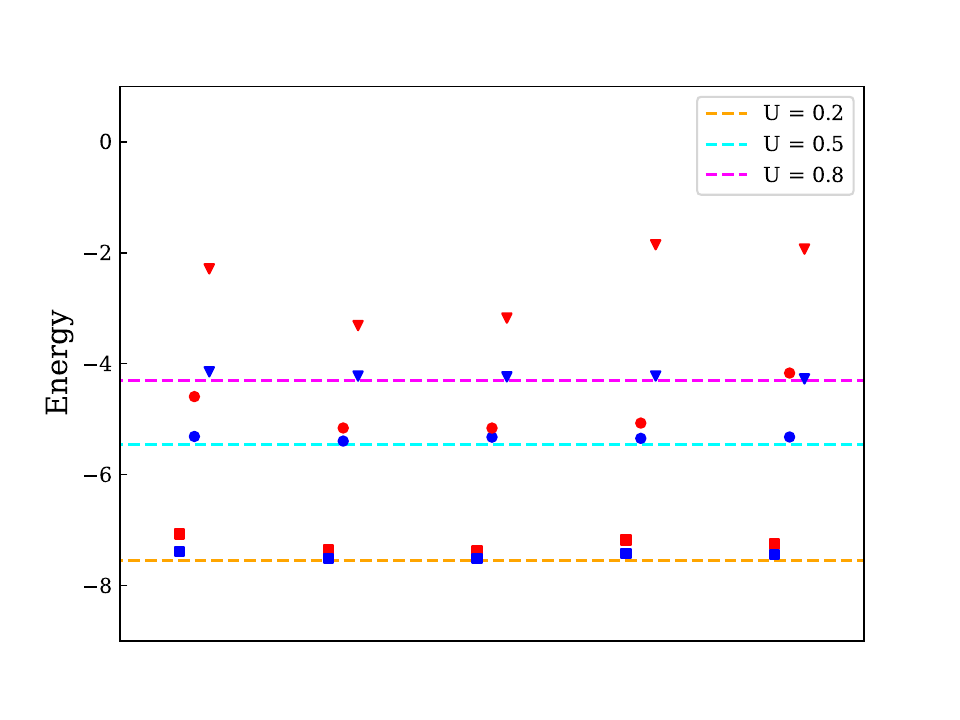 }
\caption{Five ground state energies using wave functions computed in a single
run on ibm\_brisbane with (blue symbols) and without (red symbols) readout error 
correction using the QUAT circuit. 
Triangles, circles and squares correspond to 
$U$ = 0.2, 0.5 and 0.8 respectively.
The ideal hardware error-free results are shown by dashed lines } 
\label{plt5u}
\end{figure}

\FloatBarrier

\begin{figure}[thp]
\centering
\includegraphics[scale=0.6,trim={0cm 0cm 0.5cm 0.cm},clip=true,angle=0]{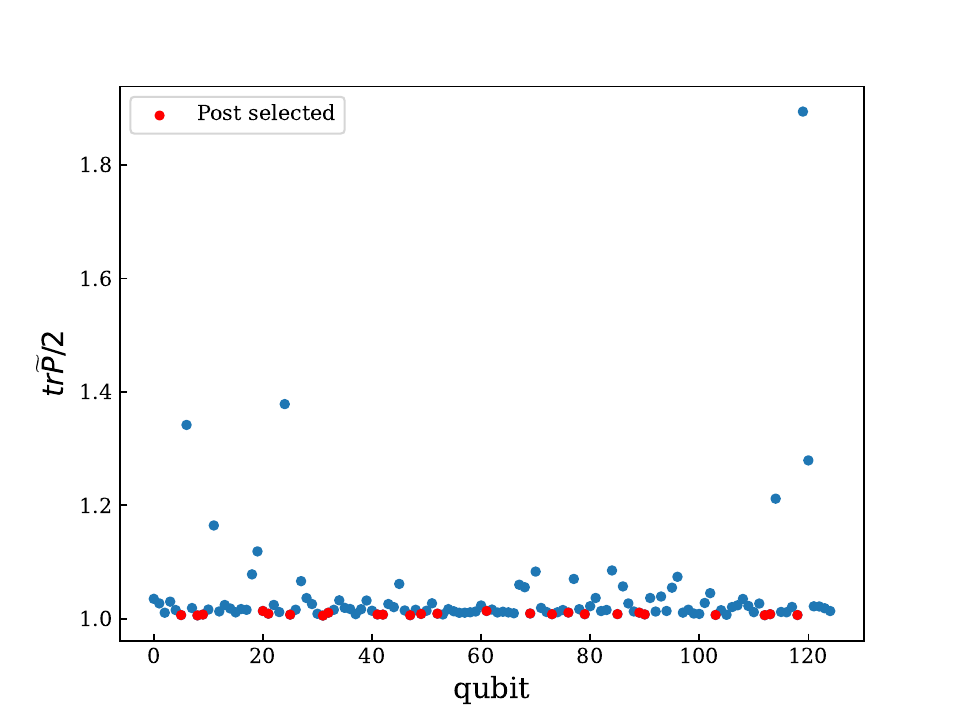 }
\caption{The readout figure of 
merit $\textrm{tr}\widetilde{P}/2$ (see Appendix A) for 125 qubits (blue symbols). The
selected qubits are shown in red. } 
\label{selc}
\end{figure}

The final step is to calculate energies on a real quantum device.
This was done on IBM Quantum resources using the Qiskit toolkit \cite{Qiskit}. 

The ground state wave function with the reduced Hamiltonian contains 26 terms.
Since normalization of the wave function provides one constraint, 25 coefficients
need to be calculated. This is done using 25 qubits each one getting a different
Fock state configuration as input and computing the corresponding wave function
coefficient. Since the ibm\_brisbane processor has 127 qubits five wave functions can
be computed at once. The five resulting ground state energies,
using 20,000 shots per measurement and without any error mitigation, 
are plotted with red symbols in Figs. \ref{plt5c}  and \ref{plt5u}. The ideal 
hardware error-free results are shown by dashed lines.

In most quantum computations, two-qubit gates are the dominant contributors to 
hardware noise. Since such gates are absent in this calculation one might 
expect readout errors to be the main source of the discrepancy between the 
red symbols in Figs. \ref{plt5c} and \ref{plt5u} and the error-free values. 
How to estimate readout error
correction factors is discussed in Appendix A. Calibration runs were done and 
applying correction factors gives results shown by the blue symbols in 
Figs. \ref{plt5c} and \ref{plt5u}.

It is possible to do better. Fig. \ref{selc} shows an example of the readout figure of 
merit $\textrm{tr}\widetilde{P}/2$ (see Appendix A) for all 125 qubits. A value
closer to one means smaller readout error. Having a measurement of each wave 
function coefficient on five qubits one can postselect the coefficient 
calculated on the qubit having the best figure of merit. The selected qubits 
are shown in red symbols. The results using postselected qubits with and without
readout error correction are shown in Table 4. As expected, with postselection
the readout error correction is quite small.

\begin{table}
\begin{centering}
\begin{tabular}{|c|c|c|c|}
\hline 
\multicolumn{1}{|c}{} & \multicolumn{1}{c}{} & \multicolumn{1}{c}{Compressed} & \tabularnewline
\hline 
U & Error-free & Uncorrected & Corrected\tabularnewline
\hline 
2 & -7.5474 & -7.3815 & -7.3697\tabularnewline
\hline 
5 & -5.4603 & -5.3665 & -5.3864\tabularnewline
\hline 
8 & -4.3665 & -4.1936 & -4.3167\tabularnewline
\hline 
\multicolumn{1}{|c}{} & \multicolumn{1}{c}{} & \multicolumn{1}{c}{QUAT} & \tabularnewline
\hline 
 & Error-free & Uncorrected & Corrected\tabularnewline
\hline 
2 & -7.5461 & -7.4059 & -7.3962\tabularnewline
\hline 
5 & -5.4600 & -5.3425 & -5.3741\tabularnewline
\hline 
8 & -4.3111 & -4.1242 & -4.2697\tabularnewline
\hline 
\end{tabular}\caption{Ground state energy using post selected qubits without and with readout
error correction applied. The value computed using a simulator without
hardware error is also shown. }
\par\end{centering}
\end{table}

It is probably possible to do even better by preselecting the qubits on which 
to calculate the 25 coefficients. This is left as an exercise for the reader.
 
\section{Summary}

Single-qubit parametrized circuits were used for constructing the variational 
wave function of a one-dimensional Bose-Hubbard model. The circuits, which
feature data re-uploading, were originally proposed for use classification 
problems \cite{PerezSalinas2020datareuploading,PhysRevA.104.012405}. 
It is shown here, that just as variational quantum states are constructed with
neural networks, one-qubit circuits can also serve as variational
\emph{Ansatze} \cite{doi:10.1126/science.aag2302}.

The circuits were trained by minimizing the ground state energy using an
ideal quantum simulator implemented in Pennylane \cite{bergholm2022pennylane}.
The results obtained were comparable, but slightly less accurate than those 
from a neural network variational wave function utilizing about 50 times more 
parameters. 

After training, the circuits were run on ibm\_brisbane. Only readout error
mitigation was applied. Typically the results were within a couple percent of 
the error-free values. For future work one may try to remove the remaining
discrepancy by a careful calibration and preselection of qubits and, 
for example, zero-noise extrapolation \cite{temme2017error} with 
unitary folding \cite{giurgicatiron2020digital} to deal with 
gate errors\footnote{Of course, the ultimate in error mitigation would be to 
use an error-correcting logical qubit. Since, in principle, only one is 
required this may be feasible in the not-too-distant future.}.

\newpage

\section*{Acknowledgment}

Thanks to Olivia Di Matteo for helpful comments.

We acknowledge the use of IBM Quantum resources in carrying out this work.
The views expressed are those of the author, and do not reflect 
the official policy or position of IBM or the IBM Quantum team

TRIUMF receives federal funding via a contribution agreement with
the National Research Council of Canada.

\section*{Appendix A}

A brief description of the readout error correction is given here.
Let $p_{ij}$ denote the probability that a qubit in state $j=\{0,1\}$
is recorded to be in state $i.$ Then 
\[
N^{O}=\left(\begin{array}{c}
N_{0}^{O}\\
N_{1}^{O}
\end{array}\right)=\left(\begin{array}{cc}
p_{00} & p_{01}\\
p_{10} & p_{11}
\end{array}\right)\left(\begin{array}{c}
N_{0}^{T}\\
N_{1}^{T}
\end{array}\right)=PN^{T}
\]
 where $N^{T}$ and $N^{O}$ denote the number of true and observed
values respectively in a set of $N_{0}+N_{1}$ measurements. The inverse
of $P,$ which will be denoted as $\widetilde{P},$ can then be used to
map observed values to true values.

To estimate $P,$ two calibration runs are done. One to prepare the
qubit in state $\left|0\right\rangle $ and measure and the other
to prepare $\left|1\right\rangle $ and measure. In this work, the
calibration runs consisted of 20,000 shots, the same as the data-taking
runs. All 125 qubits used for data-taking were calibrated in the same
run.

Generically, the diagonal elements of $\widetilde{P}$ are greater than
or equal to 1. A value of 1 would indicate that there is no readout
error. Therefore, $\textrm{tr}\widetilde{P}/2$ is used as figure of merit
for selecting qubits with small readout error.

\section*{Appendix B}

The wave function of the Bose-Hubbard model has only real coefficients
but for other models, complex coefficients may appear. To deal with
this possibility neural networks with two linear outputs $(x_{0},x_{1})$
are used. The wave function weight is then taken to be $e^{x_{0}+ix_{1}},
$\emph{i.e.}, one output provides the magnitude and the other one, the phase
\cite{saito2017, saitokato,PhysRevResearch.5.043084}. 

To generate complex coefficients with the single-qubit \emph{Ansatz} one
can prepare the circuit twice for each basis state. For example, once
for the measurement of the expectation value $\left\langle \sigma_{z}\right\rangle $
and then for $\left\langle \sigma_{x}\right\rangle .$ Then for the
magnitude of the wave function coefficient use $(1+\left\langle \sigma_{z}\right\rangle )/2$
with $e^{i\pi\left\langle \sigma_{x}\right\rangle }$ as the phase\footnote{In practice one should 
test other combinations of measurements to optimize convergence.}.

To test a variational calculation with a complex wave function a deformation
of the Bose-Hubbard Hamiltonian in the reduced basis was made by introducing
phases $e^{\pm i\varphi}$in the hopping terms preserving hermiticity.
With $t=1,U=5$ and $\varphi=\pi/2$, exact diagonalization of the Hamiltonian
gives a ground state energy of ${-4.6590}$ while a variational calculation
with a neural network using the same two-hidden-layer architecture
as in Sect. 2 (but with two outputs), gave an energy of -4.6583. 

One qubit circuits were used as described above. Convergence was
considerably slower compared to the case with of a real wave function
described in Sect. 3. To overcome this, the number of layers was increased
to 8 and the number of update steps was increased. The resulting
energies were -4.6431 for the compressed scheme circuit but only -4.5598
for the QUAT circuit. There may be scope for further optimization
but it was not pursued here.

\bibliographystyle{h-physrev4}
\bibliography{thebib}

\end{document}